\documentclass[manuscript]{aastex} 

\usepackage{emulateapj5}
\usepackage{natbib} 
\usepackage{graphicx}        
\begin{document}
\title{On the Method to Infer an Atmosphere on a Tidally-Locked Super Earth
Exoplanet and Upper limits to GJ 876d} \author{S. Seager}
\affil{Dept. of Earth, Atmospheric and Planetary Sciences, Dept. of
Physics, Massachusetts Institute of Technology, 77 Massachusetts Ave.,
Cambridge, MA, 02139.}  \author{D. Deming} \affil{Planetary Systems
Branch, Code 693, NASA/Goddard Space Flight Center Greenbelt, MD
20771}

\begin{abstract}

We develop a method to infer or rule out the presence of an
atmosphere on a tidally-locked hot super Earth. The
question of atmosphere retention is a fundamental one, especially for
planets orbiting M stars due to the star's long-duration active phase
and corresponding potential for stellar-induced planetary atmospheric
escape and erosion. Tidally-locked planets with no atmosphere are
expected to show a Lambertian-like thermal phase curve, causing the
combined light of the planet-star system to vary with planet orbital
phase.

We report {\it Spitzer} 8$\mu$m IRAC observations of GJ\,876 taken
over 32 continuous hours and reaching a relative photometric precision
of $3.9 \times 10^{-4}$ per point for 25.6~s time sampling. This
translates to a 3-$\sigma$ limit of $5.13 \times 10^{-5}$ on a planet
thermal phase curve amplitude. Despite the almost photon-noise-limited
data, we are unable to conclusively infer the presence of an
atmosphere or rule one out on the non-transiting short-period super
Earth GJ\,876d. The limiting factor in our observations was the
miniscule, monotonic photometric variation of the slightly active host
M star, because the partial sine-wave due to the planet has a
component in common with the stellar linear trend.  The proposed
method is nevertheless very promising for transiting hot super Earths
with the {\it James Webb Space Telescope} and is critical for
establishing observational constraints for atmospheric escape.

\end{abstract}

\section{Introduction}

Super Earths are a recently discovered class of exoplanets, loosely
defined to be 10~$M_{\oplus}$ or less. Since these planets have low
masses, they likely consist substantially of rocky material, making
them the first analogs of terrestrial planets in our solar system.
The search for super Earths around G through M stars continues
\citep[e.g.,][]{bagl2003, boru2003, butl2004, bonf2007, endl2008,
nutz2008, mayo2008, mayo2009}, with many discoveries of transiting super Earths
\citep{roua2009} anticipated in the next few years.

A fundamental question about rocky planets is whether or not they have
an atmosphere. The atmosphere can regulate the temperature and protect
the surface from harmful solar or cosmic rays, issues central to
surface habitability.  Extreme ultraviolet (EUV) radiation, winds, and
coronal mass ejections \citep[CMEs; ][]{lamm2007} from the host star
are the main mechanisms for driving atmospheric escape.
M stars have a much longer active phase compared to solar-like stars,
during which a star's EUV, winds, and CMEs are strong \citep{scal2007,
west2008}.  The issue of atmospheric retention, therefore, is
especially relevant for super Earths in short-period orbits around M
stars \citep[e.g.,][]{lamm2007}.

We are motivated to observationally determine whether or not a
tidally-locked close-in super Earth orbiting an M star has an
atmosphere.  The goal is to observe the planet and star in combined
light and search for thermal phase variations peaking at the
substellar point, indicative of a bare rock planet with no
atmosphere. 

The idea to study transiting hot super Earths in this way
was briefly mentioned in \citet{sels2004} and \citet{nutz2008}. On a
related topic, \citet{gaid2004} and \citet{sels2004} developed the
same idea for non-transiting terrestrial planets orbiting at the
Earth's semi-major axis from a sun-like star. The idea would be to use
a Terrestrial Planet Finder/Darwin-type interferometer
\citep[e.g.,][and references therein]{cock2009, laws2008} to null out
the star light to directly image the planet over the course of its
orbit.  \citet{gaid2004} explored light curves of planets with and
without atmospheres and oceans, for a variety of obliquities,
eccentricies, and viewing geometries. Combined light phase curves of
about half a dozen hot Jupiter transiting planets have been obtained
with the {\it Spitzer Space Telescope} \citep[e.g.,][]{harr2006,
cowa2007, knut2007, knut2009}. Even though our first proposal to
observe a hot super Earth phase curve was in 2005, three attempts at
observations were needed to reach a reasonable S/N and to mitigate
instrument systematics.
 
We begin in \S2 with a description of the motivation and background to
detect an atmosphere on a tidally-locked super-Earth. In \S3 we
describe {\it Spitzer} IRAC observations of the planet-hosting M dwarf
star GJ~876A and report on upper limits to inferring an atmosphere on
the short-period super Earth GJ~876d. In \S4 we discuss prospects for
the method to infer or rule out an atmosphere on tidally-locked
super-Earths, including transiting exoplanets.

\section{Motivation and Background}

The motivating factor for observations to discriminate between a bare
rock and a planet with an atmosphere is simply that basic atmospheric
escape estimates and calculations are inconclusive because they are
highly dependent on assumed planetary paramets (such as mass,
primordial atmosphere mass, and the past EUV radiation history of the
host star).

Atmospheric escape is a complex process thought to be driven by
nonthermal escape processes or by EUV radiative heating. The high
temperatures and consequent expanding upper atmosphere enables thermal
escape of gases. To illustrate the main uncertainties in EUV heating,
we use the energy-limited escape framework (Lecavelier des Etangs
2007).  The gravitational potential energy of the planet [J] is
\begin{equation}
\label{eq:Ep}
E_p = - \frac{G M_p m_{atm}}{\beta R_p},
\end{equation}
where G is the gravitational constant, $M_p$ is the planet mass, $R_p$
is the planet radius, and $\beta R_p$ is the radial extent of the
planet exosphere. Here $m_{atm}$ is the mass of the planet atmosphere,
which we assume to be low enough to ignore the radial atmosphere structure
in our estimate.  The
incident EUV power [J s$^{-1}$] incident on the planet is
\begin{equation}
\label{eq:power}
P_{UV} = \pi R_p^2 F_{EUV}.
\end{equation}
We assume here that the radius where the EUV energy is absorbed is
close enough to the planetary radius.  Taking the ratio of
equations~(\ref{eq:Ep}) and (\ref{eq:power}), and an
active-phase average
EUV flux $<F_{EUV}>$ leads to the lifetime of the initial planet
atmosphere
\begin{equation}
\label{eq:estlifetime}
\tau \sim \left[\frac{G}{\pi}\right]\left( \frac
{M_p}{R_p^3}\right)\frac{m_{atm}}{\beta \epsilon <F_{EUV}>}.
\end{equation}
Energy-limited escape assumes that some fraction, $\epsilon$ of the
incident UV flux heats the atmosphere and drives escape. Note that
$\beta > 1$ and $\epsilon < 1$.  This equation is valid provided that
the upper atmosphere is heated strongly enough for the relevant
elements (e.g., H, or C) to escape.

We can see the main uncertainties in atmospheric thermal escape from
equation~(\ref{eq:estlifetime}), even though the equation is
approximate at best. First, if a planet is transiting we know the
planetary density. If the planet is not transiting and we assume the
planet is solid, the density varies by less than a factor of five for
reasonable interior compositions (Mercury-like to a ``dirty water''
planet) and for a range of planetary masses \citep[e.g.,][]{seag2007}.
More uncertain is the initial outgassed planetary atmosphere mass,
which could vary widely, from less than 1 percent of the planet's
total mass up through 5 percent and even as high as 20 percent for an
initially water-rich planet \citep{elki2008}. An equally large unknown
is the incident EUV flux. All stars have very high EUV flux during
their so-called ``saturation phase'', where levels can reach thousands
of times present-day solar EUV levels. The high EUV is most critical
for planets orbiting M stars: M star saturation phases can last
1 to 3 billion years for early M stars and 6 to 8 billion years for
late M stars \citep[M5 to M8 respectively;][]{west2008}. If the
initial planetary atmosphere has not survived during the star's
saturation phase, a new atmosphere could develop in the star's quiet
phase if the outgassing rate is higher than the escape rate.

The basic idea for observations is that a tidally-locked planet with
no atmosphere, i.e. a bare rock, will show a thermal phase variation
that can be observed in the combined light of the planet-star
system. We assume short-period super Earths are tidally locked based
on dynamical timescales \citep{gold1966}.

As long as the temperature is high enough, the absorbed stellar
radiation is approximately instantaneously reradiated on a bare rock
planet. We can see this in a first order way by comparing the
radiative flux to the conductive flux. The radiative flux is,
\begin{equation}
\label{eq:FR}
F_R = \sigma T^4,
\end{equation}
where $\sigma$ is the radiation constant $\sigma = 5.670 \times 10^{-8}$~J~K$^{-4}$~m$^{-2}$~s$^{-1}$. 
Conduction is in general very inefficient in rocky material.  Heat
conduction is defined by
\begin{equation}
\label{eq:FC}
F_C = \frac{dQ}{dt dA} = -k \frac{dT}{dz},
\end{equation}
where $Q$ is heat\footnote{$Q$ is heat in the form of kinetic energy
of the motion of atoms and molecules.}, $A$ is the cross-sectional
heating area, $dQ/dt dA$ is the heat flux, $k$ in units
W~m$^{-1}$K$^{-1}$ is the thermal conductivity.  For Earth's surface
materials, $k \sim 3$. Conduction describes how heat is transfered
from a hotter to a colder region, as indicated by the negative
temperature gradient in the above equation.  

We can estimate the conductive flux by adopting different temperature
gradients $dT/dz$.  A tidally-locked planet has a maximum temperature
gradient across the surface created by the drop off of incident
stellar flux away from the substellar point. This drop off follows $(1
- \cos \theta^{1/4})$, where $\theta=0$ is defined at the substellar
point and $\theta$ is the angle away from the substellar point surface
normal, and using equation~(\ref{eq:FR}).  The temperature change
across 10~m, 100~m, or even 1~km is much less than a fraction of a
percent. The same line of reasoning can be applied to regions away
from the substellar point, also resulting in the dominance of
radiation over conduction. Hence the conductive flux across the
surface is essentially negligible, and radiation back to space will
dominate over conduction away from the heated surface element.

We can also estimate the conductive flux from the surface down into
the planet interior. We can conservatively take Earth's subsurface
temperature; a more massive planet is likely to be hotter in the
subsurface due to more heat-generating radioactive decay, and tidal
heating may also play a role. The Earth's subsurface temperature at a
depth of 500~m is about 293~K. This is deep enough so that the primary
heat source is radioactive decay of elements in the Earth's
interior. If we take GJ~876's substellar temperature of $\sim$650~K,
then, using equations~(\ref{eq:FR}) and (\ref{eq:FC}), the radiative
flux is 5000 times greater than the conductive flux, showing that
radiation back to space will dominate over conduction of energy down
into the planetary interior.  Even if we take an artificially extreme
case of a 10~K m$^{-1}$ temperature gradient just beneath the
planet's surface, the radiative flux is still a few
hundred times greater than the conductive flux. We note that Earth is
hotter in the interior than at the surface and has a temperature
gradient of about 0.02 to 0.03~K m$^{-1}$; on Earth there is no
conduction of absorbed sunlight down into the crust because conduction
only operates from hotter to colder regions.  

In summary, as long as the temperature is high enough, reradiation
will dominate over conduction. We may say that if a hot super Earth is
hot enough for its atmosphere to have escaped, it should also be in a
regime where reradiation dominates over conduction. \citet{gaid2004}
have used a different line of reasoning to argue that the thermal
inertia of even cooler rocky planets is negligible.

For instantaneous absorption and reradiation, the planetary thermal
phase curve for a tidally-locked bare rock takes the same expression
as that of a Lambert sphere \citep{sobo1975},
\begin{equation}
\label{eq:Lphasecurve}
\Phi_{\alpha} = \frac{1}{\pi} 
\left[ \sin\alpha + (\pi-\alpha) \cos \alpha \right].
\end{equation}
The phase angle, $\alpha$, is the star-planet-observer
angle. With this definition, $\alpha =
0^{\circ}$ corresponds to ``full phase'', $\alpha > 170^{\circ}$
corresponds to a thin crescent phase, and the planet is not at all
illuminated at $\alpha = 180^{\circ}$. The phase angle in terms of inclination is
\begin{equation}
\cos \alpha = - \sin i \sin 2 \pi \phi,
\end{equation}
where $\phi$ is the orbital phase.
For  the phase-dependent thermal infrared flux ratio of a bare rock we have an expression similar to that of reflected light \citep[e.g.,][]{sobo1975, char1999} with the
geometric albedo $A_g$ replaced by $(1-A_g)$,
\begin{equation}
\label{eq:Lambert}
\frac{F_p(\phi, i)}{F_*} = \left ( 1 - A_g \right) \left(\frac{R_p}{a}\right)^2
\frac{1}{\pi} \left[ \sin\alpha + (\pi-\alpha) \cos \alpha \right].
\end{equation}

 We are therefore looking for a sinusoidal variation like a
Lambert sphere, depending on the inclination of the planet.  Detection
of a Lambertian thermal phase curve is a necessary but not sufficient
condition to identify a tidally-locked exoplanet without an
atmosphere. Some tidally-locked exoplanet atmospheres, including those
with thin atmospheres or with strong absorbers at high altitudes in
thick atmospheres will shows a thermal phase curve that peaks on the
planetary day side. The only way to discriminate between a
tidally-locked bare rock exoplanet and tidally-locked planet with
a thin or thick atmosphere is that atmospheric winds likely shift the
planetary hot spot away from the substellar point.  The data needs to
be good enough to pinpoint the hottest point on the planetary day side
to see whether or not it is consistent with the substellar point.

\section{Observations and Data Analysis}

We observed GJ\, 876d, because at the time of our first proposed
observations (2005) GJ\, 876d was the only short-period super Earth
candidate available for the method to infer the presence of
an atmosphere. GJ\,876 \citep{rive2005} is a dM4 star with
$T_{eff} = 3350 \pm 300$~K \citep{mane2007}. At a distance of 4.69 pc,
GJ\,876 is a nearby star and very bright in the near-IR ($K$ = 5.0).

The planet GJ\,876d has $M \sin i = 5.89 \pm 0.54 M_{\oplus}$, $P =
1.93776 \pm 7 \times 10^{-5}$, and a semi-major axis of 0.02~AU
\citep{rive2005}. A non-transiting planet, GJ\,876d's orbital
inclination is unknown. Despite being a relatively slow rotator
\citep[96.7-day period;][]{rive2005}, and relatively inactive, GJ\,876
has a low level brightness variability of order 0.05 magnitude at
visible wavelengths \citep{shan2006}.

We observed GJ\,876 at $8\mu$m wavelength using {\it Spitzer}/IRAC.
The observations were nearly continuous for 32 hours, starting on July
16, 2008 at 01:50 UTC, using an IER in two segments.  We acquired a
total of 4204 data cubes in IRAC subarray mode, each data cube
comprising 64 exposures of 0.4-sec duration, and 32x32 pixel spatial
extent.  IRAC observations at $8\mu$m are subject to an increasing
sensitivity as the detector is exposed to light, an effect termed the
``ramp'' \citep{demi2006}.  We used a ``pre-flash'' strategy to flood
the detector with high flux levels and thus force the ramp to its
asymptotic value prior to observing GJ\,876: we observed the compact
HII region WC89 for 59 minutes immediately prior to observing GJ\,876.

Our analysis used the Basic Calibrated Data produced by version
S18.0.2 of the {\it Spitzer} analysis pipeline.  We applied the
corrections recommended by the {\it Spitzer Science Center} (SSC) to
correct for variations in pixel solid angle and flat-fielding for
point sources.  We corrected the effect of energetic particle hits by
applying a 5-point median filter to the 64-frame time history of each
pixel within each data cube.  We set the threshold of the median
filter to equal 6\% of the peak value in the average stellar image,
and we corrected discrepant pixels to the median value. We performed
aperture photometry on each of the 64 frames within each data cube,
using a circular aperture with a radius of 4.25-pixels. We centered
the aperture for each frame using a parabolic fit to the brightest 3
pixels in the X- and Y-profiles of the star, each profile derived by
summing over the orthogonal coordinate. We varied the aperture radius
to minimize the scatter in the final photometric time series. We
measured a background value using an annulus with inner and outer
radii of 6 and 12 pixels.  Each photometric point (Figure 1) was
constructed by averaging the frames within each data cube, omitting
the 1st and 58th frame \citep[see][]{harr2007}.  We also omitted
frames that were internally discrepant by more than $4\sigma$ from the
other frames within each data cube. We subtracted an average
background value for the entire time series (not frame-by-frame)
because this minimized the scatter in the time series photometry.

Figure 1 (top panel) shows the photometric time series.  We found that
a very small ramp effect (about 0.1\%, not illustrated on Figure 1)
was still present for the first $\sim$ 2 hours of the data.  We simply
dropped the first 159 minutes of data from the analysis, keeping the
last 3828 data cubes (orbit phases $\ge 0.3$).  

The most prominent feature on Figure 1 is a 2-millimag stellar flare
that occurs near phase $0.355$.  Flares are quite common on active
M-dwarfs \citep{reid2005}, although none have been previously reported
on GJ\,876.  However, three facts indicate that this is a real flare:
1) {\it Spitzer} photometry for solar-type stars has a proven
stability \citep[e.g.,][]{knut2007}, with no instrumental effects of
this type, 2) the event has the characteristic rapid rise and
exponential decline characteristic of M-dwarf flares, and 3) Knutson
et al. (2007, and private communication) observed a similar event on
the M-dwarf companion to HD\,189733.  We fit and remove a flare model
from the photometry.  The flare model has a linear rise and
exponential decline, requiring 4 parameters (rise rate, time of peak,
amplitude, and decay rate) and we fit for the minimum $\chi^{2}$ using
a gradient-expansion algorithm. We emphasize that the flare is
not a significant impediment to our analysis, because it has a well
defined shape and occupies only a relatively small fraction of the
light curve.

In addition to the flare, we observe a gradual decline in brightness
that is equivalent to 0.6 millimag per day.  GJ\,876 is known to be
variable \citep{Weis1994}, and ground-based photometry in the Johnson
B band by P. Sada (private communication) for times bracketing our
{\it Spitzer} observations shows a decrease in brightness at a rate of
$2.0\pm0.2$ millimag per day. This is part of a 97-day periodicity due
to stellar rotation, derived by \citet{rive2005}, and confirmed by
Sada's photometry concurrent with our observations.  After removing
the flare, we fit and remove the corresponding section of a 97-day
sine wave, with amplitude scaled down to fit our time series data.
Note that we expect more muted changes due to stellar rotation at this
long IR wavelength as compared to visible wavelengths.  Based on
temperatures for starspots in active late-type stars (Strassmeier et
al. 1992), Sada's 2.0 millimag per day at Johnson B scales to 0.8
millimag per day at 8$\mu$m, in good agreement with our observed
slope. This gradual decline in stellar brightness is a more
serious problem than the stellar flare for putting limits on the
planet signal, as explained in the next section.

The middle panel of Figure 1 shows the data after removal of the
stellar and linear baseline in units of the average stellar
brightness.  The scatter per point is $3.9 \times 10^{-4}$, only
slightly larger than the photon noise ($3.54 \times 10^{-4}$).  We
thus obtain about 90\% of the photon-limited signal-to-noise ratio.
The distribution of photometric values (Figure 1, middle panel) after
removal of stellar effects is indistinguishable from Gaussian.  Under
our hypothesis of a planet lacking an atmosphere, the flux modulation
due to the planet should peak at phase 0.5, computed from the
ephemeris values given by \citet{rive2005}, and updated by Rivera
and Laughlin (2008, private communication).  However, we point out
that removal of the baseline slope will affect the extraction of a
planet signal, since the partial sine-wave due to the planet has a
component in common with a linear baseline.  To account for this
effect, we fit and remove a straight line from the planet model, and
we fit the difference curve (a ``reduced sine'') to the data in order to
estimate the best-fit planet amplitude and error.

A least-squares fit of a reduced sine wave peaking at phase 0.5 gives
an amplitude of $1.26\pm1.71 \times 10^{-5}$, i.e., zero within the
errors.  To determine the error, we fit similar reduced sine waves to
100,000 bootstrap Monte-Carlo trials based on permutations of the
Figure 1 (middle panel) data.  Tabulating the distribution of
sine-wave amplitudes from these trials, we find an excellent fit to a
Gaussian distribution, with a $3\sigma$ limit = $5.13 \times 10^{-5}$.
From the photometric calibration of our data, we find the flux from
the star to be 864 mJy, so our $3\sigma$ limit on the amplitude due to
the planet is $44 \mu$Jy. At a 99\% confidence level, the flux
modulation limit due to the planet is $34 \mu$Jy. To illustrate the
$3-\sigma$ limit in a ``chi-by-eye'' fashion, the lower panel shows a
sine wave with $44 \mu$Jy amplitude in comparison to our data binned
into 8 phase intervals, as well as the nominal best-fit amplitude (not
a detection).

\section{Results}

Our Spitzer 32-hour dataset shows no evidence for the sinusoidal
variation peaking at phase 0.5. To interpret the observed limit on
flux modulation, we need to account for the orbit inclination of the
planet and the albedo. We assume a tidally-locked rotation with no
heat redistribution, because we are testing the no-atmosphere
hypothesis.  At each assumed inclination value we compute the thermal
radiance curve of the planet, emitting as a Lambertian sphere. Given
an assumed inclination, the Doppler data yield a mass, and we compute
a radius for various compositions using the mass-radius relationships
calculated by \citet{seag2007}. This information is sufficient to
define the amplitude of the planet's thermal emission curve at all
values of orbit inclination and albedo (see
equation~(\ref{eq:Lambert})).

A given upper limit on the planet thermal phase curve corresponds to a
locus of values in the inclination-albedo space illustrated in Figure
2. Tidally-locked atmosphereless planets with albedos and
inclinations in the regions below the curves in Figure 2 would have
high enough thermal phase curve amplitudes to be detected, and are
therefore ruled out by our observations. In other words, if, for a
given composition GJ~876d has an albedo and inclination such that it
lies below an upper limit curve in Figure~2, we could infer the
presence of an atmosphere (implying efficient heat
redistribution between the tidally-locked planet's day and night
side). We emphasize again that the observations actually rule out
combinations of inclination, albedo, and {\it radius}. By knowing
GJ~876d's mass at a given inclination, we can translate a mass and
adopted interior composition to a radius, using planet interior
models.

We are able to make robust (99.8\% confidence level) upper limit
statements that are inclination-, albedo-, and composition-dependent,
corresponding to regions below the curves in Figure~2.  As an example,
we would infer the presence of an atmosphere if GJ~876d is a pure
water ice planet that has an albedo less than 0.35 if the planet's
orbital inclination is 50 degrees \citep[the inferred inclination
from][]{rive2005}. We would infer the presence of an atmosphere for a
planet with the same composition and an albedo less than 0.79 and an
inclination of 84 degrees \citep[the inferred inclination
from][]{bene2002}.  A pure water ice planet is not a particularly
useful example because planets are expected to have some rocky
component.  We can also infer the presence of an atmosphere if GJ~876d
is less dense than a planet with Earth's bulk composition
(approximately 32.5\% iron core and 67.5\% silicate perovskite mantle)
and has a geometric albedo less than 0.25 and an inclination of 84
degrees. Because a bare rocky planet likely has an albedo lower than 0.25,
we would infer the presence of an atmosphere on GJ~876d for this
composition and inclination.  For other albedo/inclination values for
which we could infer the presence of an atmosphere for a given
hypothetical composition of GJ~876d, see the curves in in Figure 2. We
are unable to make conclusive statements about planets more dense than
those with Earth's composition, because size decreases with increasing
density, and smaller sized planets would have a smaller thermal phase
variation, rendering them undetectable in our data.

\section{Discussion}

We reported nearly continuous 32 hours of Spitzer IRAC data of
GJ876d. The goal was to infer whether or not GJ~876d has an
atmosphere, because a tidally-locked planet with no atmosphere is
expected to show a Lambertian-like thermal phase curve.  We are unable
to either conclusively infer the presence of an atmosphere or rule one
out. We can present inclination- and albedo-dependent statements of
what kind of planet atmospheres are ruled out if GJ876d is a planet of
relatively low average density for a solid planet (see Figure 2).

The limiting factor in our data analysis is variation of the stellar
flux. The presence of a slow linear drift increases the size of an
atmosphereless planet that can hide in the data. Even though GJ 876d
is a quiet M star, it still shows variability on a rotational time
scale of 96.7 days \citep{rive2005}. 

The serendipitous detection of a flare (Figure~1) is further evidence
for activity on a very slightly active star.  We emphasize that
the flare itself should not impede efforts to extract the planet
signal, because the flare has a specific shape and a relatively short
duration.  The detection of infrared microflares on M stars may
provide a new way to study M star activity not available from the
ground.

Ground-based visible photometric monitoring of the star during the IR
space-based measurements, combined with the rotation rate of the star
and a star spot model (area and temperature of spots) would help to
subtract any stellar variability in the light curve. Additionally,
space-based IR monitoring before and after the planet observations
would extend the data baseline to help precisely define the stellar
variability. Future observations should also use continuous
observations of the planet-star system, not just for maximizing S/N,
but also to avoid the detrimental ``ramp'' effects that would be
caused by observations of other targets of random
brightnesses. Monitoring of a planet over two or more orbital periods
would also help disentangle a linear stellar variability component.

Higher S/N would also mitigate effects of stellar variability. Our
upper limit was degraded because of overlap between a component of the
planet's variation and the stellar variability.  We fit to the data
with a ``degraded template''.  Having more photons would help to get
more precision, and tighter limits than using such a degraded template.
We caution that with more photons we might encounter another aspect of
stellar variability at a lower level. Our observations may be the
first to indicate the difficulty even low levels of stellar variability
have on observational studies in the combined-light of the planet and
star, for any kind of exoplanets orbiting an M star.

Knowing the inclination of GJ\, 876d would help in interpreting our
data. \citet{rive2005} provide an inclination for the GJ\,876 3-planet
system of $\sim$~50 degrees, based on stability arguments for the two
$\sim$Jupiter-mass planets GJ~876b and GJ~876c, assuming that the
lower mass GJ 876d is also coplanar.  {\it Hubble Space Telescope}
Fine Guidance Sensor observations \citep{bene2002}, in contrast to the
theoretical simulations, give an inclination of GJ 876b of 84$\pm6$
degrees. Even with improved observations and theoretical simulations,
the inclination of GJ 876d may remain unknown, unless we adopt the
dynamical assumption that it should be co-planar with its more
massive planet siblings.

Transiting planets provide a much better opportunity for observations
to infer whether or not an atmosphere exists from a thermal
phase curve compared to a planet with unknown orbital inclination.
Because a transiting planet's orbital inclination is near 90 degrees,
the planet-star flux contrast is maximized for a bare rock
planet. The radius of a transiting planet is known, enabling a
specification of the albedo without resort to mass-radius interior
models. (While non-transiting planets with known inclination would
still help with interpreting the results, the unknown radius remains a
complicating factor.).  We can use Figure~2 to understand what kind
of limits one could put for comparable data for a transiting
planet. If GJ 876d were a transiting planet at 90 degrees inclination
with no evident thermal phase variation, we would infer an atmosphere
on GJ~876d for almost any interior composition. The exception is
planets denser than a planet with a Mercury-like interior composition
(approximately 70\% iron core and 30\% silicate mantle); slightly
higher S/N data would be needed. For a planet with a Mercury-like
composition, the albedo upper limit is 0.15. Most bare rocky bodies in
our solar system have albedos lower than this value Mercury's
geometric albedo is 0.1 and the Moon's is 0.12). One exception is Io,
Jupiter's atmosphereless moon, with a high geometric albedo
\citep[0.63][]{cox2000} due to fresh solid sulfur deposits from
episodic volcanic outbursts, themselves caused by tidal friction with
Jupiter. Although many of the icy satellites of the outer planets also
have very high geometric albedos \citep[0.63][]{cox2000}, an
ice-covered surface is not expected for hot planets like GJ 876d and
others at very close planet-star separations. Data with S/N $\sim$3
times higher than our data set reported here would rule out such high
albedos even for the smallest plausible planets.

We have discussed how to infer the presence of an atmosphere on an
exoplanet. Can an atmosphere be ruled out based on detection of a
thermal phase curve? In principle a planet with a thick atmosphere
having inefficient heat redistribution \citep[e.g.,][]{harr2006} could produce
a signal similar to a bare rock. Measuring the phase of the orbit
variation to high precision is needed; a bare rock must peak very
close to phase 0.5, but a planet with a thick atmosphere should have a
detectable phase shift of the thermal maximum, even for relatively
inefficient heat transport. Discriminating between a planet with a
very thin atmosphere and a bare rock may be more difficult, and
requires further study.

Observation of the combined light of the planet and star as a function
of phase may also reveal the presence of a relatively strong planetary
magnetic field. For short-period planets, the planet and star magnetic
fields could interact to produce a phase curve peaking at phases near
180 degrees with a period equal to that of the planet's orbit, as has
been detected for a few hot Jupiters \citep{shko2005, dona2008,
shko2008}. The question of super Earth magnetic fields is a serious
one. Without a protective magnetic field the atmosphere might be
further eroded by stellar wind and high energy particles. A detection
of a relatively strong magnetic field would indicate the presence of a
partially liquid interior, aiding interpretation of interior
composition (and even further interesting for short-period planets
under strong tidal interaction.)

We note that Mercury is not tidally-locked to the sun, (i.e., in a 1:1
resonance). Instead Mercury, is in a 3:2 resonance, rotating three
times for every two orbits about the sun. Models show a low
probability for a planet on a circular orbit to become trapped in the
3:2 spin-orbit resonance compared to the 1:1 resonance, unless it goes
through a chaotic evolution of an eccentric orbit as Mercury likely
has \citep{corr2004}. Planets on circular orbits are therefore more
certainly tidally-locked than eccentric planets which may have had the
opportunity to be captured into higher order spin-orbit resonances.

The {\it James Webb Space Telescope}, with its 58 times more
collecting area than {\it Spitzer} will be able to make high S/N
observations of short-period super Earths. M stars are good targets,
because they are bright at IR wavelengths and the planet-star radius
ratio is high. 

In conclusion, inferring an atmosphere on a short-period,
tidally-locked super Earth that lacks a thermal phase curve is
possible for a transiting planet of any interior composition and
albedo, provided that the data set has higher S/N by about a factor of
3 than ours. During the {\it JWST} era we anticipate the birth of
studies of atmosphereless bodies and the concomitant understanding of
atmospheric escape.

\acknowledgements{We thank Feng Tian, Andrew West, Brad Hager, Leslie
Rogers, and Lindy Elkins-Tanton for many useful discussions. We thank
Eugenio Rivera and Greg Laughlin for providing an unpublished updated
ephemeris for GJ~876d, Pedro Sada for communicating his GJ\,876
photometry in advance of publication, and Heather Knutson for showing
us her unpublished photometry of the M-dwarf companion to 189733. We
thank an anonymous reviewer for a careful read of our paper. We thank
the {\it Spitzer} Science Center staff for their efficient scheduling
of our observations and for assistance in finding the best pre-flash
source. This work is based on observations made with the {\it Spitzer
Space Telescope}, which is operated by the Jet Propulsion Laboratory,
California Institute of Technology under a contract with NASA. Support
for this work was provided by NASA through an award issued by
JPL/Caltech.}

\bibliography{planets}

\begin{figure}[ht]
\centering
\includegraphics[scale=.65]{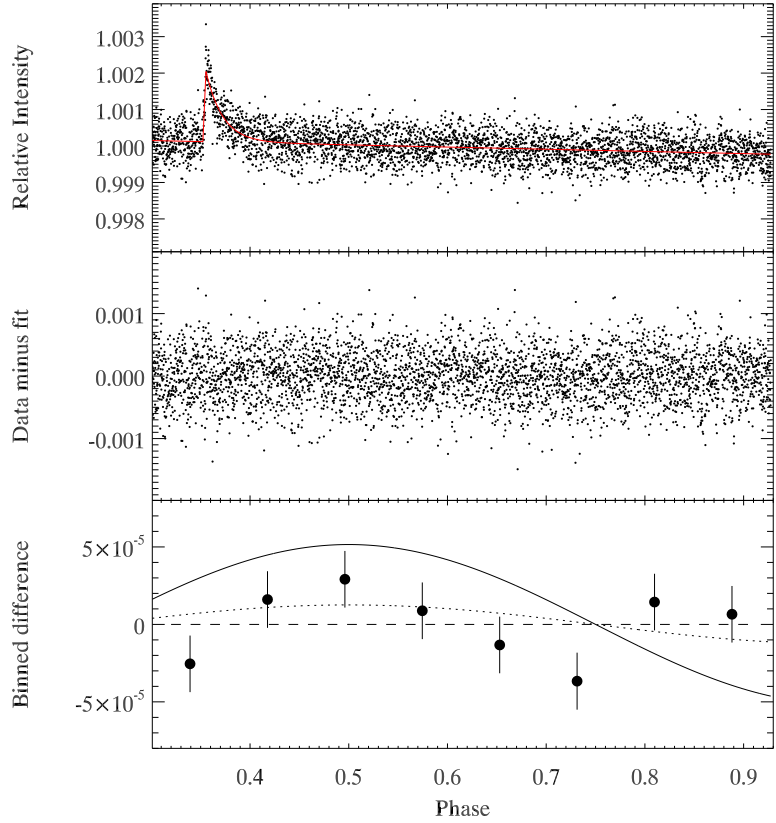}
\vspace{0.5in}
\caption{{\it Upper panel:} photometry of GJ\,876 for 3828 data cubes,
each representing 62 images.  The initial small detector ramp,
represented by 376 data points (159 minutes) was omitted from the
analysis and is not shown.  The red line is a fit of a linear (0.002
per day) brightness decrease of the star, plus a flare of amplitude
0.002 near phase 0.355.  {\it Middle panel:} Data from the upper panel
after the fit (red line) is subtracted.  {\it Lower panel:} Residuals
from the middle panel binned into 8 phase intervals, and shown in
comparison to our $3\sigma$ upper limit on the flux modulation (solid
curve, amplitude = $5.13 \times 10^{-5}$, 44\,$\mu$Jy) peaking at
phase 0.5 as expected for a planet without an atmosphere to
redistribute heat.  The dotted curve is the nominal best fit (not a
detection).}
\label{fig:atm}
\end{figure}

\begin{figure}[ht]
\centering
\includegraphics[scale=.95]{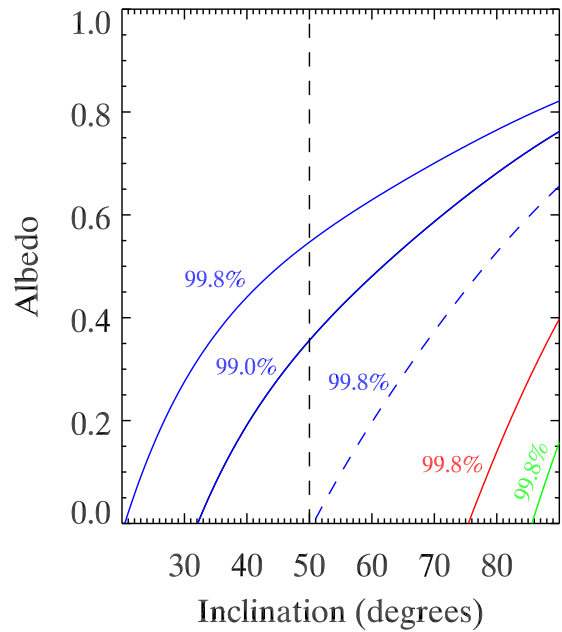}
\caption{Inclinations and albedos for which an atmosphere can be
inferred.  Each line is the locus of orbit inclination and albedo
values that would produce an 8\,$\mu$m sinusoidal flux modulation
corresponding to our 99.8\% and 99.0\% confidence upper limits (44 \&
34\,$\mu$Jy).  Color encodes planet composition: pure water ice
planets (solid blue curve); ``dirty'' ocean planets (dashed blue curve,
6.5\% by mass iron core, 48.5\% silicate mantle, and 45\% ice outer
layers, or different combinations with the same mass and radius); an
Earth-like composition (red curve, 32.5~\% iron core and 67.5\%
silicate mantle); and a Mercury-like composition (green curve, 70\%
iron core and 30\% silicate mantle). A GJ\,876 with no atmosphere is
allowed by our data above each line; the regions below the line are
forbidden at that confidence level.  In other words, if GJ\,876d has a
composition, orbit inclination, and albedo lying in the regions below
the curves, then we would infer the presence of an atmosphere.}
\label{fig:exclusion}
\end{figure}

\end{document}